# Lyman $\alpha$ Absorption and Tidal Debris

Simon L. Morris and Sidney van den Bergh
*Dominion Astrophysical Observatory,*
*National Research Council,*
*5071 West Saanich Road,*
*Victoria, B.C., V8X 4M6,*
*Canada,*
*e-mail: simon@dao.nrc.ca,*
*vandenbergh@dao.nrc.ca*

## ABSTRACT

It is suggested that a significant fraction of the low column-density absorption features seen in the spectra of quasars are produced in pressure-confined tidal debris, that was built up in small groups and clusters of galaxies over a Hubble time. We show that the space-density and cross-section of tidal tails in groups of galaxies are large enough that they could constitute a major source of the low redshift Ly$\alpha$ absorption features that are associated with galaxies. The space-density of groups within 10 Mpc of our galaxy is found to be $2.4 \times 10^{-3}$ Mpc$^{-3}$, which is close to the $\sim 6 \times 10^{-3}$ Mpc$^{-3}$ space-density calculated for Ly$\alpha$ absorbers, assuming they have a 1 Mpc radius. Other observational constraints on the properties of Ly$\alpha$ absorber such as their velocity dispersion, correlation properties, dimensions and abundances, are shown to be consistent with this hypothesis.

*Subject headings:* Galaxies:Interactions — Galaxies:Quasars:Absorption Lines

## 1. INTRODUCTION

In a pioneering paper Bahcall & Spitzer (1969) proposed that the high column-density (N(HI)$\gtrsim 10^{16}$ cm$^{-2}$) absorption lines seen at large redshifts in the spectra of quasars are produced in the halos of normal galaxies. However, they noted that the halo radii required to account for the observed number of absorption features "are an order of magnitude greater than the galactic radii normally quoted". The radii required become even larger if lower column-density absorption systems are considered. Recent investigations at low redshift, such as that by Morris et al. (1993), have shown that, for these low equivalent-width systems, the problem persists to the present day, despite the fall-off in absorber space-density with decreasing redshift. Several recent papers have attempted to explain this discrepancy. Maloney (1993) pointed out that the disk sizes measured from 21-cm radio data were probably cut off at the point where the *total* hydrogen column-density in the disk fell below a critical value of a few $\times$ 10$^{19}$ cm$^{-2}$, at which radius the gas becomes highly ionised, and the *neutral* hydrogen column-density rapidly drops below 21-cm detection limits (a few $\times 10^{18}$ cm$^{-2}$). Nevertheless, the disk cross-section at column-densities of $\gtrsim 10^{13}$ cm$^{-2}$ may be considerably larger. Hoffman et al. (1993) have extended this idea by suggesting that the *total* hydrogen column-density in disks may fall off less rapidly than an exponential in the outer regions - further expanding the possible cross-section. For disk galaxies with detected (21-cm) neutral hydrogen dimensions of $\sim$ 15 kpc, they estimate a possible radius of 250 kpc, within which N(HI)> 10$^{13}$ cm$^{-2}$.

In the present paper it is suggested instead that a significant fraction of the low column-density absorption features are, in fact, produced in tidal debris that has been built up in small bound groups and clusters of galaxies over a Hubble time.

## 2. RADII OF GROUPS OF GALAXIES

Traditionally the Local Group has been considered to consist of those galaxies that are located at distances R < 1.0 Mpc (e.g. van den Bergh 1992a). However, the work of Yahil, Tammann & Sandage (1977) suggests that somewhat more distant objects, such as IC 10, DDO 210 and the WLM system may also be gravitationally bound members of the Local Group. Their study suggests that the Local Group extends to a maximum distance $R_m \approx$ 1.6 Mpc from its center, which is located between the Galaxy and M31. Half of the (non dwarf) Local Group members are located within a projected distance of $R_h \approx$ 0.4 to 0.5 Mpc from the group center. (The exact value of $R_h$ depends on the viewing angle). For the well-studied Centaurus group (Hesser et al. 1984), which was assumed to be centered on NGC 5128 (= Cen A), it was found that $R_h$ = 8° and $R_m$ = 22°. With an assumed distance of 3.5 Mpc (van den Bergh 1992b), these values correspond to linear radii of $R_h$ = 0.5 Mpc and $R_m$= 1.5 Mpc. Such values may be fairly typical for the small groups of galaxies that dominate the field between rich clusters. We will assume that the local group is a typical small group. This assumption has been common since Hubble (1936a). In support of this, the two nearest groups (the M81 group and the South Pole group) seem quite similar in their galaxian content.

## 3. FREQUENCY OF TIDAL DEBRIS

Including the Local Group, there are 11 groups of galaxies that are know to be located within a distance of 10 Mpc (de Vaucouleurs 1975). Of these groups, three have been observed to contain large amounts of tidal debris. The total number of groups containing tidal debris may, however, be higher because not all nearby groups have yet been carefully searched for low-level 21-cm emission.

The best-known tidal feature is the Magellanic Stream in the Local Group. It is a young structure that appears to have been formed within the last few$\times 10^8$ years (Wayte 1991) by an interaction between the Large Magellanic Cloud and the Small Magellanic Cloud. Mathewson & Ford (1984) estimate the Magellanic stream to contain $1.8 \times 10^9$ M$_\odot$ of neutral gas with a column-density of $\gtrsim 10^{19}$ cm$^{-2}$. (If this gas were spread uniformly over a disk with radius 1 Mpc, it would have a surface density of $\sim 1 \times 10^{16}$ atoms cm$^{-2}$). The nearby M81 Group also exhibits extensive tidal features resulting from past interactions between NGC 3031 (M81), NGC 3034 (M82) and NGC 3077. Appleton, Davies & Stephenson (1981) estimate that $1.4 \times 10^9$ M$_\odot$ of neutral hydrogen gas is present in tails and tidal structures that are located outside the Holmberg radii of the galaxies in the M81 group. A tail-like structure with a mass of $\sim 10^9$ M$_\odot$ has been discovered in the M96 group by Schneider et al. (1983). Schneider (1985) estimates this feature to have an age of $4 \times 10^9$ yr.

In great clusters the velocity dispersion of galaxies is much larger than it is in groups. As a result one would expect only a small fraction of the interactions of galaxies in rich clusters to produce tidal tails. Furthermore hydrogen-rich spirals constitute a lower fraction of the total galaxy population in rich clusters than they do in less-populous groups (Hubble 1936b, Dressler 1980). Morris et al. (1993) claim that there is a deficit of low column-density absorbers in regions of high galaxy density. Nevertheless possible tidal features, with a mass of $\sim 1 \times 10^9$ M$_\odot$, have recently been found in the rich Virgo cluster (Giovanelli, Williams & Haynes 1991, Hoffman et al. 1993).

## 4. SPACE-DENSITY OF NEARBY GROUPS

The distance to the M66/M96 double group of galaxies has been determined using surface brightness fluctuations, planetary nebulae, globular clusters, the Tully-Fisher relation and supernovae of Type II. All of these lines of evidence yield a consistent distance modulus $(m-M)_0 = 29.9 \pm 0.1$ (van den Bergh 1992b), which corresponds to D = $9.5 \pm 0.5$ Mpc. The distance scale out to a distance of $\sim$10 Mpc is therefore independent of the still controversial numerical value of the Hubble parameter $H_0$. Including the Local Group, de Vaucouleurs (1975) lists a total of 11 nearby groups and clusters of galaxies with distances smaller than, or equal to, that of the M66/M96 group i.e. objects with D$\leq$10 Mpc. We shall combine the M66 and M96 groups (which are separated by only $8°.1$ corresponding to a projected linear separation of 1.3 Mpc) and regard them as a single group. Furthermore we shall, to avoid possible bias, omit the Local Group. The remaining 10 galaxy groups within 10 Mpc yield a space-density of $2.4 \times 10^{-3}$ Mpc$^{-3}$. Because the environs of the Local Group are located in the outermost envelope of the Virgo supercluster this density may be somewhat above the average density of small groups and clusters in nearby regions of the Universe.

This space-density of groups can be compared with that of galaxies brighter than M$^*$ (M$_{b_J}$=-19.5 for H$_0$=100 km s$^{-1}$ Mpc$^{-1}$), which can be derived from the results of Loveday et al. (1992). Integrating their fit to a Schecter luminosity function, one gets a density of $3 \times 10^{-3}$ h$^3$ Mpc$^{-3}$, with H$_0$ = 100 h km s$^{-1}$ Mpc$^{-1}$. Assuming that on average each group contains two galaxies brighter than M$^*$, and that nearly all such galaxies lie in groups, and taking h$\approx$0.75 (Jacoby et al. 1992, van den Bergh 1992b), one gets a space-density for galaxy groups of $\sim 4 \times 10^{-3}$ Mpc$^{-3}$. This value is in reasonable agreement with that obtained above. In the subsequent discussion a space-density of $2 \times 10^{-3}$ groups Mpc$^{-3}$ will be adopted.

## 5. SPACE-DENSITY OF ABSORBERS

The number of absorbers with cross-section $\sigma (= \pi R^2)$ between z=0 and z=z$'$ is given by

$$N = \left(\frac{c}{H_0}\right)\frac{\sigma n}{3q_0^2}[(1+2q_0 z')(q_0 z' - 1 + 3q_0) + (1 - 3q_0)], \quad (1)$$

where n is the number of absorbers per Mpc$^3$ (Maloney 1992). With q$_0$ = 0, z$'$ = 0.15 and R expressed in Mpc,

$$n = 6.6 \times 10^{-4} Nh/R^2. \quad (2)$$

With R $\approx$ 1 Mpc, h=0.75, and using the observations of 3C273 (Morris et al. 1991) which give $9 \lesssim N \lesssim 17$ for N(HI) $\gtrsim 10^{13}$ cm$^{-2}$, this yields n $\sim 6 \times 10^{-3}$ Mpc$^{-3}$. In view of the uncertainties in the adopted values of N, h and particularly R, this value is uncertain by a factor of a few. It is noted that the density of $\sim 6 \times 10^{-3}$ absorbers per Mpc$^3$ is close to the value $\sim 2 \times 10^{-3}$ groups per Mpc$^3$ estimated in § 4. It is therefore concluded that available data are consistent with the hypothesis that a significant fraction of Ly$\alpha$ absorbers are small groups and clusters of galaxies that contain tidal debris. In § 6., we describe evidence that only 20% of the low redshift Ly$\alpha$ absorbers are associated with galaxies. If this is correct, the group space-density estimated above would roughly match the space-density of such absorbers.

The crucial, and presently unanswerable, question is: what is the ratio of the cross-section of galaxy disks at neutral hydrogen column densities of $\gtrsim 10^{13}$ cm$^{-2}$ to the cross-section of tidal tails? A comparison at column-densities of $\gtrsim 10^{19}$ cm$^{-2}$ *is* possible. Typical observed HI disks have radii of 15 kpc, or a surface area of 700 kpc$^2$. The Magellanic stream covers a region approximately 120° $\times$ 10°. If it is at the distance of the Magellanic clouds, this yields a surface area of 1000 kpc$^2$. Thus the surface areas are comparable. Extrapolating down six orders of magnitude in column-density to estimate the relative cross-sections for Ly$\alpha$ forest absorbers would be foolish, but we note that the very effect that probably truncates the outer regions of galaxy *total*[1] hydrogen disks (i.e. tidal forces from nearby galaxies) will also probably produce tails and streams. In the local group, if they had disks of the size proposed by Hoffman et al. (1993), the Milky way and M31 total hydrogen disks would have radii 1/2 their separation, and a large number of the local group dwarf galaxies would repeatedly pass though them.

## 6. DISCUSSION

Carilli, van Gorkom & Stocke (1989) and Sargent & Steidel (1990) show convincing examples of high column-density absorption systems produced by tidal tails. In these cases, the QSO line of sight passes within $\sim 10$ $h^{-1}$ kpc of the galaxy. The question we wish to address here is whether one can already rule out an extrapolation from the observations of tidal tails at column-densities of $> 10^{18}$ cm$^{-2}$, down to the Ly$\alpha$ forest regime of $10^{13}$ cm$^{-2}$ and impact parameters from the QSO line of sight to the galaxy of $>100$ $h^{-1}$ kpc.

Apart from their space-density, there are a number of other properties of the low redshift Ly$\alpha$ absorbers which can be checked for consistency with the idea that they are produced by gas in tidal tails in groups of galaxies:

1. The velocity dispersions of the gas in the high column-density portions of tidal tails is known to be low. Mathewson & Ford (1984) quote typical FWHM of 18-25 km s$^{-1}$. The absorbers at low redshift tabulated by Morris et al. (1991) were in general only marginally resolved, and had FWHM $\lesssim 60$ km s$^{-1}$. These small velocities are inconsistent with the idea that the absorbers are gas in virial equilibrium in halos of giant galaxies (which have virial velocities of 100 - 300 km s$^{-1}$, see Mo & Morris 1993 for further discussion of this point). Tidal material in rich clusters of galaxies may be evaporated or destroyed by further interactions, although some examples are known (Giovanelli, Williams & Haynes 1991, Hoffman et al. 1993). Patterson & Thuan (1992) measure a velocity FWHM of 110 km s$^{-1}$ for the tidal material stripped from UGC 7636 in the Virgo cluster, a rather large value compared to the majority of Ly$\alpha$ absorbers, but their large beam width of 22 arcmin means that this FWHM refers to the whole extended HI tail, and is not necessarily the FWHM of any single gas clump.

2. The three dimensional shapes of the low redshift absorbers are not known. Smette et al. (1992) place a lower limit of 12 $h^{-1}$ kpc on the transverse dimensions of the high redshift absorbers. Barcons & Fabian (1987) used the distribution of high redshift absorber column densities to argue that a model with highly non-spherical gas clouds would be required unless the gas density of each cloud was very non-uniform. Hippelein & Meisenheimer (1993) claim to have actually imaged a z=3.8 Ly$\alpha$ absorber, deriving dimensions of 5×20 $h^{-1}$ kpc. This size seems rather small compared to the total lengths of present day tidal tails, but could easily be explained as a sub-clump within such a structure.

3. Material stripped from objects in which star formation has occurred (either dwarf satellite galaxies like the Magellanic clouds, or the disks of L* galaxies) should contain processed material, and hence have significant metals. For example, Mirabel, Dottori & Lutz (1992) quote a value of 8.4 for 12+log[O/H] for the dwarf galaxy at the tip of the Antennae (NGC4038/39), a factor of a few below solar. This object is claimed to have been formed from tidally stripped material. Such a value is at the lower end of the range found for disks of spiral galaxies (e.g. Oey & Kennicutt 1993), possibly because the material forming the dwarf at the tip of the Antennae tidal tail was stripped from the outer edge of one of the interacting disk systems. If low redshift Ly$\alpha$ absorbers belong to two distinct populations, one being tidal tails, and the other not associated with galaxies, then one might expect a bimodal distribution in metallicity. Because of their low column-densities, and probable high ionisation, it is very difficult to measure the metallicity of individual Ly$\alpha$ absorbers. Observations of the wavelength range containing SiIV and CIV for the 3C273 absorbers are scheduled to be obtained by HST late in 1993. At high redshifts, some progress has been made by summing up large numbers of absorbers and measuring a mean metallicity. Lu (1991) claimed a detection of CIV in this way. After making ionisation corrections, he derived a relative logarithmic carbon abundance [C/H] of -3.2 compared to solar. This value is three orders of magnitude lower than that expected for tidally stripped material, and this could be taken to indicate that tidal material must make up a very small ($\leq$ 0.1%) fraction of the high redshift absorbers. However, it should be emphasized that the ionisation correction is large and uncertain, and that the fraction of tidal tails compared to 'isolated' absorbers may be a function of look-back time (see below).

---

[1] As has been noted by Maloney (1993) and Corbelli and Salpeter (1993), galaxy *neutral* hydrogen disks, as viewed in 21 cm, are truncated by photoionisation by the extra-galactic UV background.

4. The low redshift absorbers were also found to show a significant correlation with galaxies near the line-of-sight (Morris et al. 1993). This point is explored in more detail in Mo & Morris (1993), who conclude that the 3C273 data are consistent with the absorbers being a 80%:20% mix of randomly distributed absorbers, and absorbers that are associated with cataloged galaxies. It is this latter population which we suggest could be dominated by tidal material. A testable prediction of this hypothesis is that those galaxies that are near absorbers should themselves have close galaxy neighbors. The absorber-galaxy correlation function of higher redshift Ly$\alpha$ absorbers is not known, although their weak auto-correlation function is usually taken to imply that it is small.

5. The lifetime of gas in tidal tails is generally not known. Lower limits can be derived from the observed age of known tails, which are up to a few $\times 10^9$ years (Schneider 1985). Wayte (1991) argues that the Magellanic stream is a very young structure. His conclusion is based both on the time it would take to double in width due to random bulk motion, and also its timescale for free expansion. However, essentially all of the models and discussion about the confinement of high redshift 'isolated' Ly$\alpha$ clouds can be carried over to individual 'clouds' in tidal tails. This suggests that the timescale for free expansion may only be a lower limit to the true expected lifetime of the Magellanic stream. The pressure confined models for Ly$\alpha$ absorbers (Ikeuchi & Turner 1991 and references therein) may be helped by the higher pressure environment likely to prevail in galaxy groups. Mulchaey et al. (1993) show that a nearby 'normal' group contains gas with a temperature of 1 keV and density of $5 \times 10^{-4}$ cm$^{-3}$. Also, as mentioned above, the highly non-spherical nature of tidal tails may help explain the observed column-density distribution, without having to appeal to the huge mass range normally required in this model. Gravitationally confined models may also be possible. Barnes & Hernquist (1992a,b) discuss the formation of self-gravitating clumps of material in tidal tails. One major difference from the 'CDM minihalo' model (Miralda-Escude & Rees 1993 and references therein) however, is that such clumps are predicted to have rather little 'dark matter' in them.

6. A final point to consider is the evolution of the space-density of tidal tails with time. For Ly$\alpha$ absorbers in a fixed (and low) column-density interval, rather weak evolution in number density per redshift interval is seen between z=0 and z=2.0. At higher redshifts there is, however, a fairly rapid increase in number density with increasing redshift. This evolution has been shown to be explicable within either the pressure confined (Ikeuchi and Turner 1991) or CDM minihalo models (Mo, Miralda-Escude & Rees 1993). Two factors will dominate the space-density and cross-section in neutral hydrogen of tidal tails: (1) the rate of tidal interactions, and (2) the intensity of the UV background flux ionising the optically thin gas clouds in the tails. The rate of galaxy mergers within the CDM cosmology has been estimated recently by Carlberg (1992) and by Kauffmann & White (1993). Carlberg (1992) states that the rate increases rapidly with increasing redshift. The UV background evolution has been studied by Bechtold et al. (1987) and Miralda-Escude & Ostriker (1990). Both state that the background flux was approximately constant between z=2 and z=4, and falls rapidly as one moves towards the present day. Without detailed models it is impossible to say which of these two effects will dominate. Therefore, it is not yet possible to predict whether tidal tails will be a larger or smaller fraction of the absorbers at high redshifts.